%% file: main.tex
\pdfoutput=1 % Make arXiv happy
\documentclass[conference]{IEEEtran}
\IEEEoverridecommandlockouts
\usepackage{graphicx}
\usepackage{textcomp}
\usepackage[dvipsnames]{xcolor}
\usepackage{booktabs}
\usepackage{xspace}
\usepackage{xparse}
\usepackage[inline]{enumitem}
\usepackage{tikz}
\usepackage{svrsymbols}
\usepackage{tikzpeople}
\usepackage[expansion,protrusion]{microtype}

\usepackage[colorlinks=true, urlcolor=blue,
            linkcolor=blue, anchorcolor=blue]{hyperref}
            
\usetikzlibrary{arrows.meta, calc, backgrounds, shadows.blur}

\newcommand{\website}{accompanying \href{https://github.com/lfd/qrep123}{website}\xspace}
\NewDocumentEnvironment{nop}{}{\bgroup}{\egroup}
\newcommand{\QS}{QSW\xspace}
\begin{document}

\def\break{\\}
% Title playground. 
%\title{Plug\&Play Reproducibility\break for Quantum Software Experiments}
%\title{Auto-Generate Reproducibility Packages \break for Quantum Software Experiments
\title{1-2-3 Reproducibility \break for Quantum Software Experiments % 
%
%\title{Self-Service Reproduction Packages %\break for Quantum Software Experiments % cool! evtl. "self-service"
% TODO - DOUBLE BLIND
\thanks{WM acknowledges supported from the German Federal Ministry of Education and Research (grant~13N15646, ``quantum technologies'').}
}

\author{
\IEEEauthorblockN{Wolfgang Mauerer}
\IEEEauthorblockA{\textit{Technical University of}\\
\textit{Applied Sciences Regensburg}\\
\textit{Siemens AG, Corporate Research}\\
\href{mailto:wolfgang.mauerer@othr.de}{wolfgang.mauerer@othr.de}}
\and 
\IEEEauthorblockN{Stefanie Scherzinger}
\IEEEauthorblockA{\textit{Chair of Scalable Database Systems} \\
\textit{University of Passau}\\
Passau, Germany \\
\href{mailto:stefanie.scherzinger@uni-passau.de}{stefanie.scherzinger@uni-passau.de}}
}
\maketitle
\makeatletter\def\break{\penalty-\@M}\makeatother

\begin{abstract}
Various fields of science face a reproducibility crisis. For quantum software engineering
as an emerging field, it is therefore imminent to focus on proper
reproducibility engineering from the start. Yet the provision of reproduction
packages is almost universally lacking. Actionable advice on how to build 
such packages is rare, particularly unfortunate in a field with many
contributions from researchers with backgrounds outside computer science. In this article, we argue how to rectify this
deficiency by proposing a 1-2-3~approach to reproducibility engineering for quantum
software experiments: Using a meta-generation mechanism, we generate
DOI-safe, long-term functioning and dependency-free reproduction packages. They
are designed to satisfy the requirements of professional and learned societies solely on
the basis of project-specific research artefacts (source code, measurement and
configuration data), and require little temporal investment by researchers.
Our scheme ascertains long-term traceability even when the
quantum processor itself is no longer accessible. By drastically lowering the technical
bar, we foster the proliferation of reproduction packages in quantum software
experiments and ease the inclusion of non-CS researchers entering the field.
\end{abstract}


%\begin{small}
%    \textcolor{red}{Position paper (2 pages): Describing a well-argued position in any area of quantum software analysis, evolution, and reengineering. \textbf{DOUBLE BLIND!!}}
%\end{small}

\section{Introduction}

The ACM reproducibility guidelines (v1.1)~\cite{acm_terms} consider an experiment  \emph{reproducible} when a different team with a different experimental setup  is able to confirm the published results.\footnote{The ACM has repeatedly changed the definition of their reproducibility terminology in an incompatible way; we are aware of the arising confusion.}
Despite the universally acknowledged importance of reproducibility, the vast majority of
researchers agree that we are facing a reproducibility crisis~\cite{Baker:2016,Jasny:2011}
% ,Ivie:2018
in almost all domains of science, including quantum computing 
research~\cite{DBLP:conf/icse-qse/CamposS21,DBLP:journals/corr/abs-2007-07047,
DBLP:journals/corr/abs-2108-09744,Fingerhuth2018}.

In the emerging field of quantum software (\QS) engineering, reproducibility engineering has also been recognised as a grand challenge~\cite{DBLP:conf/icse-qse/CamposS21,DBLP:journals/corr/abs-2007-07047,DBLP:journals/corr/abs-2108-09744,Fingerhuth2018}, with first contributions on managing the reproducibility of software bugs~\cite{DBLP:journals/corr/abs-2108-09744,DBLP:conf/icse-qse/CamposS21}, or on exploring the 
parameter search space for quantum optimisation tasks~\cite{DBLP:journals/corr/abs-2110-05555}. Yet so far, little actionable advice has been given on how to engineer \QS experiments for reproducibility.
As we will argue shortly, despite the general credo that reproducibility is important, at this moment in time, published articles that are accompanied by reproduction packages are rare.

Building a working \emph{reproduction package} goes beyond providing a DOI to some
repository hosting data, code, and setup instructions. 
Rather, a gold-standard reproduction package~\cite{Heil2021} bundles all research artefacts
required to conduct the experiment (such as source code, libraries, or input data), and
contains a dispatcher script that allows for executing and evaluating the experiment via a single
command.

%By now, several high profile CS conferences expect authors to submit such a reproduction package along with their article.

\paragraph*{Quantum-specific challenges} 
Building reproduction packages for \QS shares many challenges of classic software engineering, most importantly managing
complex software stacks that non-trivially interact with hardware. 

Reproducibility challenges specific to \QS experiments are
plentiful~\cite{McGeoch:17}: Quantum computing hardware is usually provided as a cloud
service, with vendor-controlled access and configurations. % controlled
%by the vendor. 
%Scientists trying to reproduce a result
Reproducers
may not have ready access to
the very same machine (especially after prolonged periods of time). Even if they
do, they are very likely to find the configuration changed, given that most quantum
hardware is still in  its infancy and subject to change even during the
operational phase.

Apart from a unique model/type specification of the
employed machine, it is necessary to provide information
on 
\begin{enumerate*}[label=(\alph*)]
\item input generation methods,
\item qbit counts,
\item connectivity topology,
\item any methods used to transform inputs,
\item approaches used to map/embed logical qbits onto physical qbits,
\item postprocessing methods and utilities employed, and
\item programming/initialisation and readout times. 
\end{enumerate*}
Quantum software experiments may involve manual tuning, which
necessitates specifying any employed heuristics/policies. Finally, details on how
runtimes are measured are important, especially in cloud settings that may
involve interference by access schedulers.

\paragraph*{Synopsis}
%We perform a quantitative review of the state-of-the-art,
%focusing on the number of publications on QS experiments which can
%be reproduced, and argue there is need for action.
We quantitatively review the state of reproducibility 
in \QS experiments, and recognise a need for action.
We then introduce a customisable  \emph{meta package template} that
generates end-to-end, one-click reproductions with adaptation examples for common
quantum hardware. It can be found at the
\website/\href{https://doi.org/10.5281/zenodo.5113867}{DOI}),
together with a video tutorial. Only little input is required
from the scientists.

\paragraph*{Vision}
The ease-of-use of our reproduction package
template %(we go so far as calling it a 1-2-3 approach) 
enables the 
proliferation of working reproduction packages for \QS experiments.
Moreover, lowering the threshold for building reproduction packages
can foster skills in \QS engineering, an area difficult to master for
newcomers~\cite{shaydulin20}.

% Mention that some of the authors are also to blame for not publishing 
% reproduction packages for some of their quantum computing papers!
% Since information exchange in quantum software mostly happens in workshops
% given the novelty of the field, we focus on these venues.

% We only consider source code linked in the paper, and did not perform a
% manual search on github and other forges.
% Reproduction package == data/code on DOI, without further long-term
%                         executability considerations
% IEEE Quantum Week: https://ieeecs-media.computer.org/media/conferences/resources/quantum-program-2021.pdf
%
\begin{table}[htb]
\caption{Reproducibility packages in recent \QS
publications.}\label{tab:stats}
\vspace*{-1em}\begin{center}\begin{tabular}{llrrrrr}
     \toprule
     Venue & Year & \# Papers & \# Exp &  \# Src & \# Repro\\
     \midrule
     \href{https://icsa-conferences.org/2021/workshops/qsa/}{QSA@ICSA} 
           & 2021 & 4 & 2 & 1 & 0\\
     \href{http://ceur-ws.org/Vol-2705/}{Q-SET@QW} & 2020 & 6 & 2 & 0 & 0\\
     \href{https://quset.github.io/qset2021/program.html}{Q-SET} & 2021 & 4 & 2 & 1 & 0\\
     \href{https://events.cels.anl.gov/event/27/timetable/?print=1&view=standard}{QCS@SAC} & 2021 & 11 & 10 & 8 & 0\\
     \href{https://apeqs.lfdr.de/2020/#/home}{APEQS@FSE} & 2020 & 4 & 1 & 0 & 0\\
     \href{https://netsys2019.org/workshops/qtop19/}{QTOP@Netsys}& 2019 & 18 & 12 & 2 & 1\\
     \href{https://q-se.github.io/qse2020/}{Q-SE@ICSE} & 2020 & 8 & 3 & 2 & 0\\
     \href{https://q-se.github.io/qse2021/}{Q-SE@ICSE} & 2021 & 8 & 3 & 1 & 1\\
     \bottomrule
\end{tabular}\end{center}\vspace*{-2em}
\end{table}

\section{State of the Art}
\emph{State of reproducibility.}
Table~\ref{tab:stats} provides an overview about the state of reproducibility
in quantum software. Based on eight workshops from 2019--2021, where we currently observe most
activities related to QSW, we reviewed 63 papers (column ``\#~Papers''; detailed
results are on the \website). %\todo{\scriptsize DOI?}
We
classified them according to whether they describe artefacts (column ``\#~Exp''; algorithmic source
code or experiments to be run on quantum hardware or simulators) that should be
provided in a reproducible form. We also state the number of papers that
 \emph{at least} provide source
code on a non-permanent software forge like GitHub (column ``\#~Src''),
and the fraction of papers that provide a reproduction package (column ``\# Repro''; we count anything provided under a
DOI-safe permanent location, even if it does not meet the standards suggested in
Ref.~\cite{acm_terms}).

The conclusion is simple: Although most published research is concerned
with executing code and simulations, only slightly more than half of the publications
provide source code \emph{at all}, and only \emph{two} papers come with artefacts on
a DOI-safe location! Clearly, there is a mismatch between desirable reproducibility
qualities as mandated by professional associations (and, therefore, the community itself),
and scientific day-to-day reality. 

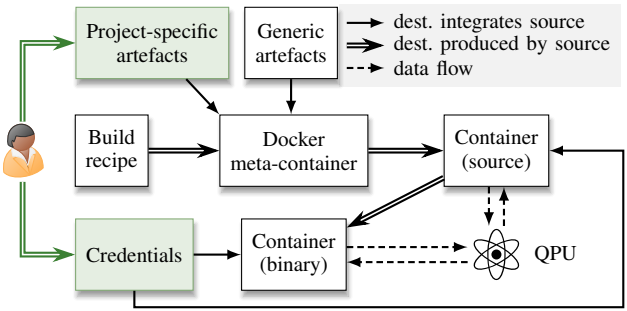
\begin{figure}[thb]
\vspace*{-0.4em}\input{qsaner_chart}\vspace*{-0.8em}

\caption{Components and workflow for 1-2-3 reproducibility engineering: The scientist provides project-specific artefacts (green). A docker meta-container then generates a reproduction package consisting of a source container
and a pre-built binary docker container. Both containers can trigger the \QS experiment, record measurement and configuration data, and analyse results. Credentials for accessing commercial quantum hardware and cloud services
are not encoded into the reproduction package, but specified as parameter.}
\label{fig:container}
\end{figure}

\section{A Template Proposal}
To make \QS experiments reproducible with little extra
effort for researchers, we suggest a generic, adaptable template
(with full source available on the \website) as reference for
\QS experiments as illustrated in Fig.~\ref{fig:container}. It includes
examples for dealing with DWave and IBMQ systems. Based on the supplied artefacts, the
generated package comprises
\begin{enumerate*}[label=(\alph*)]
\item the complete source code for the calculations performed in the paper,
\item 
results obtained on quantum computers (to ensure
traceability beyind quantum processor availability) and via classical simulation,
\item
a concise documentation of each reproduction step,
\item a ``one-click'' dispatcher to run the % computation and measurement
pipeline, 
\item ideally, means of generating the underlying research paper, including graphs
and tables. % based on reproduced results.
 \end{enumerate*}

The replication package in both, source and pre-built binary form, is intended
to be made available at a DOI-safe, long-term stable location. The former
is easier to extend by peer researchers, but assumes external components
on volatile locations are still available. The latter ensures long-term reproducibility,
since it does not have external dependencies.

To avoid impediments for adoption 
that might arise from technologies like docker---many researchers
in \QS do not have computer science backgrounds, and even if, it might not be systems-centric---, we suggest this generative approach.

% TODO: QC research often domain-driven; repro package template is low-threshold
% and eliminates need to focus on low-level technological CS details. Plug-and-play, 
% Batteries included. Boilerplate code

\section{Conclusion}
Reproducibility of experiments is key in scientific research. Yet the
adoption of methods to ensure this quality is currently sub-par in the
quantum software literature. Since the field is still in relative infancy, we propose
sustainable rectification of the situation from the start, by providing a generic scheme
that generates reproduction packages that are tailored to quantum software experiments, and that can be
used with minimal effort. 

%\printbibliography
\bibliographystyle{IEEEtran}
\bibliography{literature}
\end{document}

%% file: qsaner_chart.tex
% \usepackage{tikz}
% \usepackage[dvipsnames]{xcolor}
% \usepackage{tikzpeople}
% \usepackage{svrsymbols}

% \usetikzlibrary{arrows.meta, calc, backgrounds, shadows.blur}

\begin{tikzpicture}[
    nodes={font=\footnotesize},
    artefact/.style={draw, rectangle, align=center, inner sep=5pt, fill=white, blur shadow={shadow xshift=0.7pt}, minimum height=27pt, inner sep=4pt},
    proj artefact/.style={artefact, draw=OliveGreen!60, fill=OliveGreen!10},
    head/.style={draw, circle, minimum size=7pt, inner sep=0, draw=OliveGreen!60},
    body/.style={draw=OliveGreen!60},
    scientist/.style={alice, minimum size=1.5em}, % or person
    basic arrow/.style={-{latex}, thick},
    double arrow/.style={-{Stealth[length=3mm, width=2mm]}, double, thick},
    proj arrow/.style={double arrow, OliveGreen},
    dashed arrow/.style={basic arrow, dash pattern=on 2.5pt off 1.5pt},
    legend label/.style={},
    atom/.style={font=\Huge, inner sep=1pt},
]
    
    % nodes
    \node[scientist] (scientist) at (0,0) {};
    
    \node[proj artefact, above=0.6 of scientist, xshift=48.8] (proj-arts) {Project-specific\\artefacts};
    \node[artefact, right=0.2 of proj-arts] (gen-arts) {Generic\\artefacts};
    
    \node[artefact, right=0.42 of scientist] (build-recp) {Build\\recipe};
    \node[artefact, right=of build-recp, xshift=-1.7] (meta-cont) {Docker\\meta-container};
    \node[artefact, right=of meta-cont] (src-cont) {Container\\(source)};
    
    \node[proj artefact, below=0.55 of scientist, xshift=42] (creds) {Credentials}; 
    \node[artefact] (bin-cont) at (meta-cont |- creds) {Container\\(binary)};
    \node[atom, label=right:{QPU}] (qpu) at (src-cont |- bin-cont) {$\atom$};
    
    % edges
    \draw[proj arrow] (scientist) |- (proj-arts);
    \draw[proj arrow] (scientist)  |- (creds);
    \draw[basic arrow] (proj-arts) -- (meta-cont.north west);
    \draw[basic arrow] (gen-arts.south) -- (gen-arts.south |- meta-cont.north);
    
    \draw[double arrow] (build-recp) -- (meta-cont);
    \draw[double arrow] (meta-cont) -- (src-cont);
    
    \draw[dashed arrow] ($(src-cont.south) + (-0.1,0)$) -- ($(qpu.north) + (-0.1, 0)$);
    \draw[dashed arrow] ($(qpu.north) + (0.1,0)$) -- ($(src-cont.south) + (0.1, 0)$);
    
    \draw[basic arrow] (creds) -- (bin-cont);
    \draw[basic arrow] (creds) -- ++(0,-0.7) -- ++(6.5,0) |- (src-cont);
    \draw[double arrow] (src-cont) -- (bin-cont);
    \draw[dashed arrow] ($(bin-cont.east) + (0,0.1)$) -- ($(qpu.west) + (0,0.1)$);
    \draw[dashed arrow] ($(qpu.west) + (0,-0.1)$) -- ($(bin-cont.east) + (0,-0.1)$);
    
    % legend
    \coordinate (legend1-start) at (4.3,1.7);
    \node[legend label, right=0.5 of legend1-start] (legend1-end) {dest. integrates source};
    \draw[basic arrow] (legend1-start) -- (legend1-end);

    \coordinate (legend2-start) at ($(legend1-start) + (0,-0.3)$);
    \node[legend label, right=0.5 of legend2-start] (legend2-end) {dest. produced by source};
    \draw[double arrow] (legend2-start) -- (legend2-end);
    
    \coordinate (legend3-start) at ($(legend2-start) + (0,-0.3)$);
    \node[legend label, right=0.5 of legend3-start] (legend3-end) {data flow};
    \draw[dashed arrow] (legend3-start) -- (legend3-end);
    
    \begin{scope}[on background layer]
        \fill[gray!10] ($(legend3-start.south west) + (-0.1, -0.25)$) rectangle ($(legend1-end.north east) + (0.35, 0)$);
    \end{scope}

\end{tikzpicture}